\documentclass[twocolumn]{aastex61}
\pdfoutput=0 
\usepackage{amsmath,amstext}
\usepackage[T1]{fontenc}
\usepackage{apjfonts} 
\usepackage[figure,figure*]{hypcap}
\usepackage{booktabs}
\usepackage{rotating}

\shorttitle{Modeling Ices in Magellanic Cloud YSOs}
\shortauthors{Pauly \& Garrod}

\begin{document}

\title{Modeling CO, CO$_2$ and H$_2$O ice abundances in the envelopes of young stellar objects in the Magellanic Clouds }

\author{Tyler Pauly}
\affiliation{Cornell Center for Astrophysics and Planetary Science, Department of Astronomy, Cornell University, Ithaca, NY 14853, USA}
\author{Robin T. Garrod}
\affiliation{University of Virginia Departments of Chemistry and Astronomy, Charlottesville, VA 22904, USA}
\begin{abstract}
Massive young stellar objects in the Magellanic Clouds show infrared absorption features corresponding to significant abundances of CO, CO$_2$ and H$_2$O ice along the line of sight, with the relative abundances of these ices differing between the Magellanic Clouds and the Milky Way. CO ice is not detected towards sources in the Small Magellanic Cloud, and upper limits put its relative abundance well below sources in the Large Magellanic Cloud and the Milky Way. We use our gas-grain chemical code MAGICKAL, with multiple grain sizes and grain temperatures, and further expand it with a treatment for increased interstellar radiation field intensity to model the elevated dust temperatures observed in the MCs. We also adjust the elemental abundances used in the chemical models, guided by observations of HII regions in these metal-poor satellite galaxies. With a grid of models, we are able to reproduce the relative ice fractions observed in MC massive young stellar objects (MYSOs), indicating that metal depletion and elevated grain temperature are important drivers of the MYSO envelope ice composition.  Magellanic Cloud elemental abundances have a sub-galactic C/O ratio, increasing H$_2$O ice abundances relative to the other ices; elevated grain temperatures favor CO$_2$ production over H$_2$O and CO. The observed shortfall in CO in the Small Magellanic Cloud can be explained by a combination of reduced carbon abundance and increased grain temperatures. The models indicate that a large variation in radiation field strength is required to match the range of observed LMC abundances. CH$_3$OH abundance is found to be enhanced in low-metallicity models, providing seed material for complex organic molecule formation in the Magellanic Clouds. 
\end{abstract}


\section{Introduction}
Much of our understanding about the details of star formation comes from investigations of stars and the interstellar medium (ISM) in the galaxy, yet the peak of star formation occurred in the past at lower metallicity \citep{Madau14}. The Magellanic Clouds, local dwarf satellites of the Milky Way, provide an astronomical laboratory to study the process of star formation in a metal-poor environment. Comparison studies between sites of star formation in the Magellanic Clouds and the Milky Way can illuminate the metallicity dependence of local physical processes via observational tracers such as molecular emission and absorption features. Knowledge of multiple molecular abundances can begin to separate effects of metallicity from local physical parameters, e.g. the radiation environment and the dust temperature. 

Mid-infrared spectral observations of embedded young stellar objects (YSOs) in the Milky Way (MW) have found a wealth of solid-state features, showing high column densities of ices such as H$_2$O, CO, CO$_2$, and CH$_3$OH \citep{Gerakines99,Gibb04}. H$_2$O is the most abundant ice, with a typical column density of order 10$^{-4}$ with respect to total hydrogen; CO$_2$ is next, at an average value of CO$_2$:H$_2$O $\simeq$ 0.2 \citep{Boogert04}. CO and CH$_3$OH ices follow at lower abundance, though with nearly an order of magnitude of variation between lines of sight. These ices are found in the dense, cold envelopes surrounding the luminous central source, and they hold information on the collapse history of the progenitor dense molecular cloud via e.g.~the polar to apolar ratio of the CO and CO$_2$ ice features \citep{Gibb00}. They are processed to some extent by the internal radiation source, yet a complete explanation for the variation in observed galactic YSO ice abundances is not in hand. Local environment likely plays a role, with changes in the nearby interstellar radiation field (ISRF) or the cosmic ray ionization rate affecting gas and grain surface chemistry. Additionally, variations in the underlying elemental abundances of the collapsing cloud will influence the general chemistry and total ice column density. 

Observations of massive YSOs (MYSOs) in the nearby Magellanic Clouds show a marked difference in ice abundances with respect to galactic counterparts  \citep{vanLoon05,Oliveira09,Oliveira11,Oliveira13,Shimonishi08,Shimonishi10,Shimonishi16a}. \citet{Shimonishi10} and \citet{Oliveira11} have detected H$_2$O, CO and CO$_2$ ice in massive YSOs in the Large Magellanic Cloud (LMC); they found bulk compositional differences in LMC sources compared to their galactic counterparts, shown in elevated CO$_2$ ice or depleted H$_2$O ice, with an average value for CO$_2$:H$_2$O of 0.32. \citet{Oliveira11,Oliveira13} found only an upper limit for CO ice in all Small Magellanic Cloud (SMC) sources studied, with abundances (with respect to their H$_2$O columns) a factor of three to ten lower than their galactic counterparts. \citet{Oliveira11} and \citet{Shimonishi16a} provided additional near-infrared spectra of a sample of LMC MYSOs, with detections or upper limits for CH$_3$OH ice towards all sources studied.

In addition to ice abundance variations, the properties of gas and dust in the Magellanic Clouds also differ from their galactic counterparts. A significant fraction of molecular gas in galaxies like the metal-poor Magellanic Clouds reside in a CO-dark phase, where an extended photodissociation region keeps all atoms but hydrogen in atomic form \citep{Madden12,Madden16,RomanDuval14}. LMC dust temperatures are elevated; \citet{Bernard08} used {\it Spitzer Space Telescope} data to find a globally-averaged value of 21.4 K, or 23 K in the 30 Dor region. They also performed spectral energy distribution (SED) fitting with a variable ISRF, finding that increasing ISRF strength by a factor of $\sim$ 2.1 best fits average LMC observations. \citet{Galametz13} analyzed data from {\it Spitzer, Herschel} and the {\it Large Apex Bolometer Camera} to better model the sub-millimeter component of the dust SED. Their best-fit dust temperature for the N158-N159-N160 region of the LMC is 27 K.

Dust temperatures in the SMC have been measured towards H II regions and YSOs. Towards N27, a bright H II region in the SMC bar, \citet{Caldwell97} finds dust temperatures of 33-40 K, while \citet{Heikkila99} finds a similar range of 35-40 K. \citet{vanLoon10} use observations of YSOs in the Magellanic Clouds to find dust temperatures of 37-51 K in the SMC versus 32-44 K in the LMC. \citet{Chiar98} finds dust and ice temperatures in galactic YSO counterparts to be generally less than 30 K, with some measurements of 23-25 K.

We lack detailed measurements on the ISRF of the Magellanic Clouds; apart from the ISRF fitting of \citet{Bernard08} in the LMC, \citet{Vangioni80} and \citet{Pradhan11} provide evidence for a factor of 4 to 10 increase in the UV and far-UV field strength in the SMC when compared to the solar neighborhood.

Chemical models by \citet{Garrod11} found that dust temperatures can strongly affect the abundances of key grain surface molecules. Above dust temperatures of $\sim$ 12 K, grain surface diffusion of CO becomes rapid, and the reaction CO + OH $\rightarrow$ CO$_2$ + H efficiently produces CO$_2$. The authors also presented a gas-grain model with free-fall collapse which reproduced the threshold visual extinctions for detection of H$_2$O, CO$_2$ and CO ices. In this work, we will utilize a similar approach for an investigation of ice abundances towards YSOs in the low metallicity environments of the LMC and SMC.

Past work by \citet{Acharyya15, Acharyya16} showed that for chemical models with reduced elemental abundances, reproducing observed CO$_2$/H$_2$O ice abundance ratios towards Magellanic Cloud YSOs requires models with A$\mathrm{_V}$ = 10 and warm dust temperatures, either 20 K $\leq$ T$_d$ $\leq$ 25 K or 35 K $\leq$ T$_d$ $\leq$ 45 K. These works produce CO$_2$ via mobile CO in CO + OH; below 20 K, immobile CO causes CO$_2$/H$_2$O to drop below 0.01. They used static cloud models, keeping A$\mathrm{_V}$ and density fixed while running an array of models across a range of dust temperatures. They did not include the mechanism introduced by \citet{Garrod11} for CO$_2$ production, whereby surface O + H $\rightarrow$ OH can proceed atop a CO ice surface, and the energy of OH formation overcomes the modest activation energy barrier of the CO + OH $\rightarrow$ CO$_2$ + H reaction.

\begin{figure}
\label{fig:obs}
\includegraphics[width=\columnwidth]{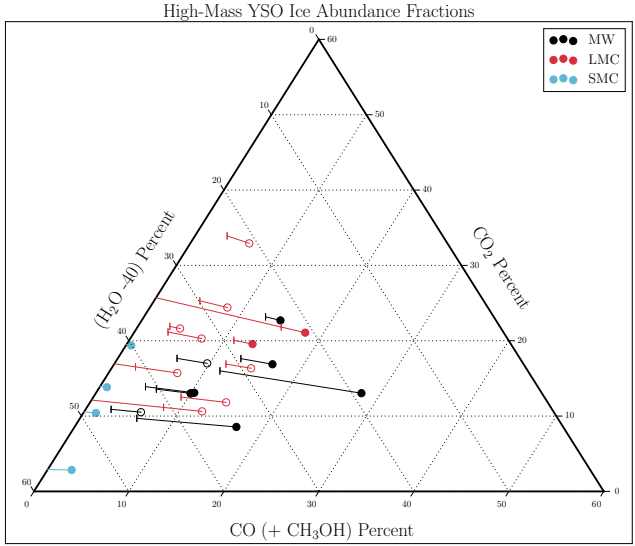}
\caption{The relative abundances of the four major ices in massive young stellar objects in the Milky Way (black), LMC (red), and SMC (blue). SMC sources have only upper limits on CO and no information on CH$_3$OH; SMC points show the composition at the upper limit value with a line drawn to zero CO abundance. For LMC and Milky Way sources, the vertical tick shows the H$_2$O:CO:CO$_2$ composition, while the circles show the composition including CH$_3$OH ice as H$_2$O:(CO+CH$_3$OH):CO$_2$. For LMC sources with only an upper limit on CO, lines have been drawn to zero CO abundance, while sources with an upper limit on CH$_3$OH ice have an open circle. \citep[Ternary figure style from][]{Harper15}}
\end{figure}

We collate observations of MYSOs for which H$_2$O, CO, CO$_2$ and CH$_3$OH detections or upper limits are available, excluding CH$_3$OH for SMC sources (toward which no measurements of CH$_3$OH have yet been achieved). Table~\ref{table:obstbl} lists the total sample we will use for model comparison. Figure~\ref{fig:obs} shows the observations from Table~\ref{table:obstbl} in a ternary H$_2$O:CO$_2$:CO ice diagram. The ternary plot describes the relative abundances of this three-component ice system. Importantly, we also consider methanol (CH$_3$OH) ice, a key component for galactic YSOs and now detected toward some LMC MYSOs. To include this fourth component on a ternary diagram, we include a second point for those sources with methanol detections or upper limits; these points show the fractional abundance of H$_2$O:CO$_2$:(CO+CH$_3$OH). This pairing choice of (CO+CH$_3$OH) is chemically motivated, as CH$_3$OH is primarily formed from the successive hydrogenation of CO on grain surfaces \citep{Watanabe02,Watanabe03,Watanabe04,Fuchs09,Cuppen09}. The figure shows a transition in composition, with some blending between some LMC and galactic sources.

Using the single-point free-fall collapse model detailed by \citet{Garrod11} and \citet{Pauly16}, we investigate parameters responsible for the chemical variation amongst MYSOs in the galaxy, LMC and SMC. We take the elemental abundances and dust temperatures to be the parameters of interest for the model study. We describe our model methods and parametrization in \S 2; results of the model grid are shown in \S 3; discussion of the results and additional parameters of interest are presented in \S 4; \S 5 concludes the study with some thoughts on future work.

\begin{table}[ht!]
\begin{center}

\begin{tabular}{r l c c c c}
\toprule
 & Source & H$_2$O & CO & CO$_2$ & CH$_3$OH\\
\midrule
MW & Mon R2 IRS 2$^{ac}$ & 77.1\% & 5.8 & 13.0 & 4.1 \\
 & RAFGL989$^{ad}$ & 62.7 & 12.6 & 22.7 & 2.0 \\
 & RAFGL2136$^{acd}$ & 76.5 & 4.5 & 13.1 & 5.9 \\
 & RAFGL7009S$^{ade}$ & 59.0 & 9.5 & 13.0 & 18.5 \\
 & W33 A$^{acd}$ & 74.2 & 5.4 & 8.6 & 11.8 \\
 & NGC 7538 IRS1$^{ae}$ & 73.3 & 6.0 & 17.0 & \textless 3.7 \\
 & NGC 7538 IRS9$^{acd}$ & 66.5 & 12.4 & 16.9 & 4.2 \\
 & W3 IRS 5$^{ade}$ & 83.7 & 2.4 & 10.5 & \textless 3.4 \\
\midrule
LMC & ST1$^b$ & 69.0 & 11.4 & 16.4 & \textless 3.3 \\
 & ST2$^b$ & 77.2 & \textless 2.1 & 15.7 & \textless 5.0 \\
 & ST3$^b$ & 73.8 & 3.2 & 21.7 & \textless 1.3 \\
 & ST4$^b$ & 67.5 & 4.6 & 24.4 & \textless 3.5 \\
 & ST5$^b$ & 72.2 & 3.4 & 20.3 & \textless 4.1 \\
 & ST6$^b$ & 60.8 & \textless 14.7 & 21.1 & 3.4 \\
 & ST7$^b$ & 60.8 & 3.3 & 32.9 & \textless 2.9 \\
 & ST10$^b$ & 67.2 & 10.7 & 19.6 & 2.5 \\
 & ST14$^b$ & 73.9 & 8.7 & 11.8 & \textless 5.6 \\
 & ST16$^b$ & 76.9 & \textless 7.8 & 10.6 & \textless 4.7 \\
\midrule
SMC & IRAS 00430\---7326$^f$ & 88.3 & \textless 1.3 & 10.4 & ... \\
 & S3MC 00540\---7321$^f$ & 85.4 & \textless 0.8 & 13.8 & ... \\
 & S3MC 00541\---7319$^f$ & 80.0 & \textless 0.6 & 19.4 & ... \\
 & IRAS 01042\---7215$^f$ & 94.6 & \textless 2.6 & 2.8 & ... \\
 \bottomrule
\end{tabular}
\caption{Fractional ice columns for observed high-mass young stellar objects in the Milky Way, Large Magellanic Cloud and Small Magellanic Cloud. The abundance of each species is shown relative to the sum of the four column densities, in percent.\label{table:obstbl}$^a$: \citet{Gibb04}, $^b$: \citet{Shimonishi16a}, $^c$: \citet{Brooke99}, $^d$:  \citet{Boogert08}, $^e$: \citet{Dartois99}, $^f$: \citet{Oliveira13}}
\end{center}
\end{table}

\section{Methods}
We use the gas-grain chemical code MAGICKAL and its associated chemical network, first presented by \citet{Garrod13} and updated by \citet{Pauly16} to include a grain-size distribution consisting of five grains. The model features 475 gas-phase species and 200 grain surface species with a network of roughly 9000 reactions and processes. Grain surface species are tracked in two separate phases, surface and mantle; the surface species participate in desorption, reaction and diffusion across grain sites, while the ice mantle is treated as a separate phase that is coupled to the surface. Bulk diffusion in the mantle ice is treated explicitly, allowing reactions within the mantle, as well as exchange between surface and mantle components; however, for the low temperatures involved in this work we treat the mantle phase as inert except for the transfer of surface material into the bulk, as the mantle grows. The model uses the modified-rate approach detailed in \citet{Garrod08} (method "C") to account for possible stochastic effects in the surface chemistry. The chemical network also includes photodissociation and photoionization processes, with photons sourced either from the ambient field or the cosmic ray-induced UV field.

\subsection{Physical Model}
The updated MAGICKAL code utilizes a grain size distribution. Following \citet{Pauly16}, we adopt the power-law fit to the size distribution of silicate grains in the ISM provided by \citet{Mathis77}, which follows the relationship $dn/da= C a^{-3.5}$. Upper and lower limits to the distribution adopted in the model, as well as the power law constant are given in Table~\ref{table:params}. The upper limit from \citet{Mathis77} is loosely constrained by extinction curve measurements, while the lower limit is a practical modeling constraint imposed by stochastic single-photon heating of very small dust grains \citep{Cuppen06}. At sizes smaller than roughly $\sim 0.02 \mathrm{\mu m}$, grains experience single photon heating to temperatures sufficient to evaporate surface species at time scales shorter than accretion rates; therefore, they are not expected to contribute significantly to ice-mantle formation. The power law constant is taken from \citet{Draine84}, though it is scaled down to match the original gas-to-dust ratio; this is required due to our shift in a$\mathrm{_{min}}$ and a$\mathrm{_{max}}$ from the values given by \citet{Mathis77}.

We assume a spherical shape for grains, with the cross-sectional area as $\sigma = \pi a^2$. We discretize the grain size distribution into five bins, equally spaced in log($\mathrm{\sigma}$). For each bin, i, the mean cross-sectional area of grains in the bin, $\langle \sigma_i \rangle$, is calculated via the power law. This $\sigma$ and its associated radius are used as representative values for all grains in that bin.

\begin{table}[ht!]
\begin{center}
\begin{tabular}{l l}
\toprule
Parameters & Values\\
\midrule
Initial n$\mathrm{_{H}}$ & 3 $\times 10^3$ $\mathrm{cm^{-3}}$ \\
Final n$\mathrm{_{H}}$ & 2 $\times 10^4$ $\mathrm{cm^{-3}}$ \\
Initial $\mathrm{A_{V}}$ & 3.00 \\
Final $\mathrm{A_{V}}$ & 10.627 \\
Final time & 5 $\times 10^6$ yr \\
$\mathrm{T_{gas}}$ & 10 K \\
$\mathrm{a_{min}}$ & 0.02 $\mathrm{\mu m}$ \\
$\mathrm{a_{max}}$ & 1.00 $\mathrm{\mu m}$ \\
Power law constant & 4.436 $\times 10^{-26}$ $\mathrm{cm^{2.5}/H}$ \\
Cosmic ray ionization rate & 1.3 $\times 10^{-17} \mathrm{s^{-1}}$ \\
\bottomrule
\end{tabular}
\caption{Model Physical Parameters \label{table:params}}
\end{center}
\end{table}

The power law constant determines the total abundance of dust. \citet{RomanDuval14} measured the gas-to-dust ratio in the LMC, finding a range of 160 to 500 for the dense to diffuse ISM, compared to 100 to 250 for the Milky Way. We follow \citet{Acharyya15} and use a value of 175; this value is fixed for all models.

The power law exponent from \citet{Mathis77} concentrates cross-sectional area in grains with the smallest radius, which are more numerous, whereas dust mass and volume are concentrated in the largest, least-populous grains. Small grains will drive the bulk surface chemistry due to concentrated accretion cross-section.

\subsection{Collapse Method}
We use free-fall collapse to simulate the density of the YSO envelope, using the methods presented by \citet{Garrod11}, following \citet{Spitzer78} and \citet{Brown88}. The density increases following:
\begin{equation}
\frac{dn}{dt} = \left(\frac{n^4}{n_i}\right)^{1/3} \left\lbrace 24\pi G m_H n_i \left[\left(\frac{n}{n_i}\right)^{1/3} - 1 \right]\right\rbrace^{1/2}
\end{equation} 
with $\mathrm{n_i}$ the initial density, G the gravitational constant, and $\mathrm{m_H}$ the mass of a hydrogen atom. Initial and final densities and visual extinctions are given in Table~\ref{table:params}, where the final visual extinction is not a parameter but is determined from the other three parameters via the relation $A_V = A_{V,0} (n_H / n_{H,0})^{2/3}$.

\subsection{Dust Temperatures}
We model the evolution of dust temperature as a function of the visual extinction and dust radius, following methods outlined in \citet{Garrod11}. We add an additional variable in a model-dependent interstellar radiation field (ISRF). With dust heating from the ISRF equal to cooling from dust radiation, we solve: 

\begin{equation}
\int_0^{\infty} Q_{\nu} J_{\nu} D_{\nu}\left( A_V \right) d\nu = \int_0^{\infty} Q_{\nu} B_{\nu}\left(T_d\right) d\nu
\end{equation}

where $\mathrm{Q_{\nu}}$ is the frequency-dependent efficiency of absorption or emission, $\mathrm{J_{\nu}}$ is the radiation field intensity incident on the cloud edge, $\mathrm{D_{\nu}}\left(A_V\right)$ is the attenuation of the radiation field at a given frequency for a given $\mathrm{A_V}$, and $\mathrm{B_{\nu}\left(T_d\right)}$ is the Planck function. We use the assumption from \citet{Krugel03} for the right-hand side of Equation 1, expressed in cgs units, valid for grains between the small- and large-grain limits, to find:

\begin{equation}
\int_0^{\infty} Q_{\nu} J_{\nu} D_{\nu}\left( A_V \right) d\nu = 1.47 \times 10^{-6} a T_d^6
\end{equation}

with $a$ as the dust grain radius. We use tabulated data on line-of-sight extinction profiles with $\mathrm{R_V}=5$ from \citet{Cardelli89} and \citet{Mathis90} to determine $\mathrm{D_{\nu}(A_{V})}$. This approach assumes plane parallel geometry. The absorption efficiency of dust grains at wavelengths relevant to the ISRF is approximated as $\mathrm{Q_{\nu}^{abs}} \propto a \lambda^{-1.5}$ with a maximum $Q$ value of 2.0, a reasonable assumption for carbonaceous grains. Silicate dust has a more complex (and generally weaker) absorption behavior in the 0.1 - 10 $\mathrm{\mu m}$ range. Our treatment therefore implicitly considers only carbonaceous grains. Of note, we treat the growth of the ice mantle during model evolution as extra grain material and not explicitly as ice for the value of Q$\mathrm{^{abs}_{\nu}}$.

\begin{figure}[!ht]
\label{fig:td_av}
\epsscale{1.1}
\plotone{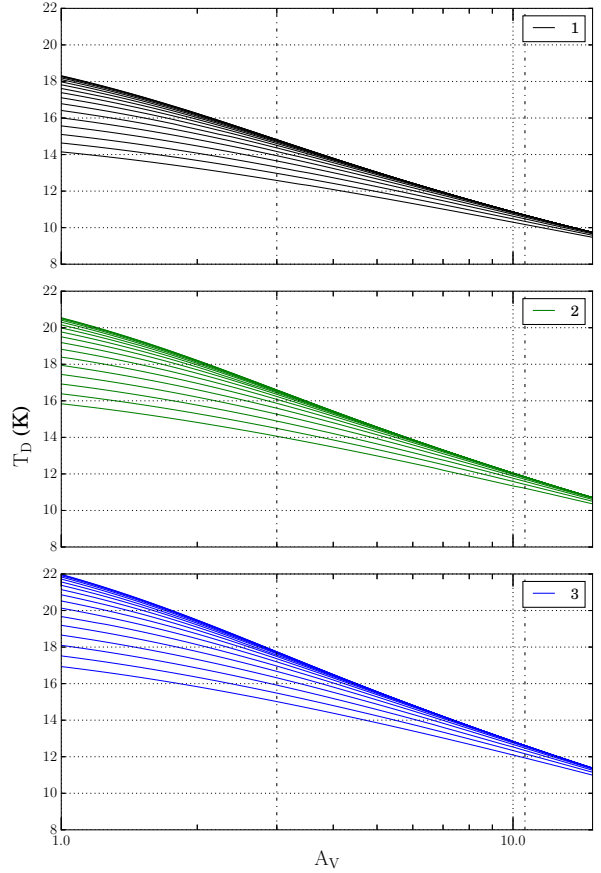}
\caption{Tracks of dust temperature versus visual extinction with lines for dust grains of constant radius, from $\mathrm{a_{lower} = 10^{-1.7} \mu m}$ to $\mathrm{a_{upper} = 10^{-0.1} \mu m}$; smaller grains have higher temperature. The top panel shows results for a stellar intensity factor of 1.0, the middle panel for 2.0, and the bottom panel for 3.0. Vertical dash-dotted lines indicate the span of $\mathrm{A_V}$ covered in our models' collapse.}
\end{figure}

We approximate the ISRF in various environments by modifying the multi-component fit from \citet{Zucconi01} for the Milky Way. The fit includes contributions from three discrete stellar black-body populations, both hot and cool diffuse dust components, and the cosmic microwave background. To simulate variation in ISRF intensity in the Magellanic Clouds, we scale the stellar components uniformly, from the base factor of 1.0 to 3.0 in increments of 0.5. The resulting dust temperatures are shown in Figure~\ref{fig:td_av} for a range of dust radii spanning the sizes explored in our models, with the smallest grains having the highest temperature. Note that the largest radius bin is 10$^{-0.1}$ $\mathrm{\mu m}$ and not 10$^0$ due to the discretization of the power law into five sizes in each model. The dashed vertical lines show the extent of $\mathrm{A_V}$ covered during the model collapse; the increase in A$\mathrm{_V}$ during the collapse process results in a general cooling and a flattening of the temperature distribution with respect to grain size. 

The dust temperature tracks in Figure~\ref{fig:td_av} are for grains of constant radius, but it should be noted that the effective grain radius is not constant during the model evolution; as gas species accrete and form an ice mantle, the grain radius grows, producing further cooling \citep[see][]{Pauly16}.

The ISRF factor used to scale the dust heating is also used to scale the photo-ionization and photo-dissociation rates in the model, as the stellar component of the ISRF is the primary source of UV photons. Ionization and dissociation via the secondary UV field from cosmic rays are treated separately.

\begin{table}[ht]
\begin{center}
\begin{tabular}{l l l l}
\toprule
Element & MW & LMC & SMC\\
\midrule
H 				& 5.000(-5) & 5.000(-5) & 5.000(-5) \\
H$_2$ 			& 0.499975 	& 0.499975 	& 0.499975 	\\
O 				& 3.200(-4) & 2.140(-4) & 1.047(-4) \\
C$\mathrm{^+}$ 	& 1.400(-4) & 6.310(-5) & 1.585(-5) \\
N 				& 7.500(-5) & 1.12(-5) 	& 2.820(-6) \\
C/O Ratio 		& 0.438		& 0.295 	& 0.151 	\\
\bottomrule
\end{tabular}
\caption{Elemental abundances, listed with respect to total hydrogen number density, $\mathrm{n_H}$. The first column represents galactic abundances, taken from \citet{Garrod11}. The second column is used as representative values for the LMC, taken from \citet{Peimbert03}. The final column is the most depleted abundances considered and are taken from the sample of SMC HII regions in \citet{Kurt98}.\label{table:elemabundances}}
\end{center}
\end{table}

\subsection{Elemental Abundances}
To model the ISM of the metal-poor galaxies, we deplete the heavy elemental abundances in the initial setup of our models. The Magellanic Clouds have bulk metallicity of $\mathrm{Z_{LMC} \sim 0.4 Z_{\odot}}$ and $\mathrm{Z_{SMC} \sim 0.2 Z_{\odot}}$ \citep{Russell92}. \citet{Kurt98} collated observations of eight LMC and six SMC HII regions with updated atomic transition data to find the mean abundances of carbon, nitrogen and oxygen. \citet{Peimbert03} collected a UV-visible spectrum of 30 Doradus in the LMC with the Very Large Telescope; with 269 identified emission lines, they calculate the total abundance of carbon, nitrogen and oxygen. The results differ slightly if recombination lines are used rather than collisionally excited lines; we have chosen to use the results of the collisional lines. These two studies provide us with ISM compositions to model the metal-depleted environments associated with star formation in the Magellanic Clouds. The abundance values are shown in Table~\ref{table:elemabundances}; these abundances will be referred to as MW, LMC, and SMC.

\section{Results}

\begin{figure*}
\epsscale{1.17}
\plotone{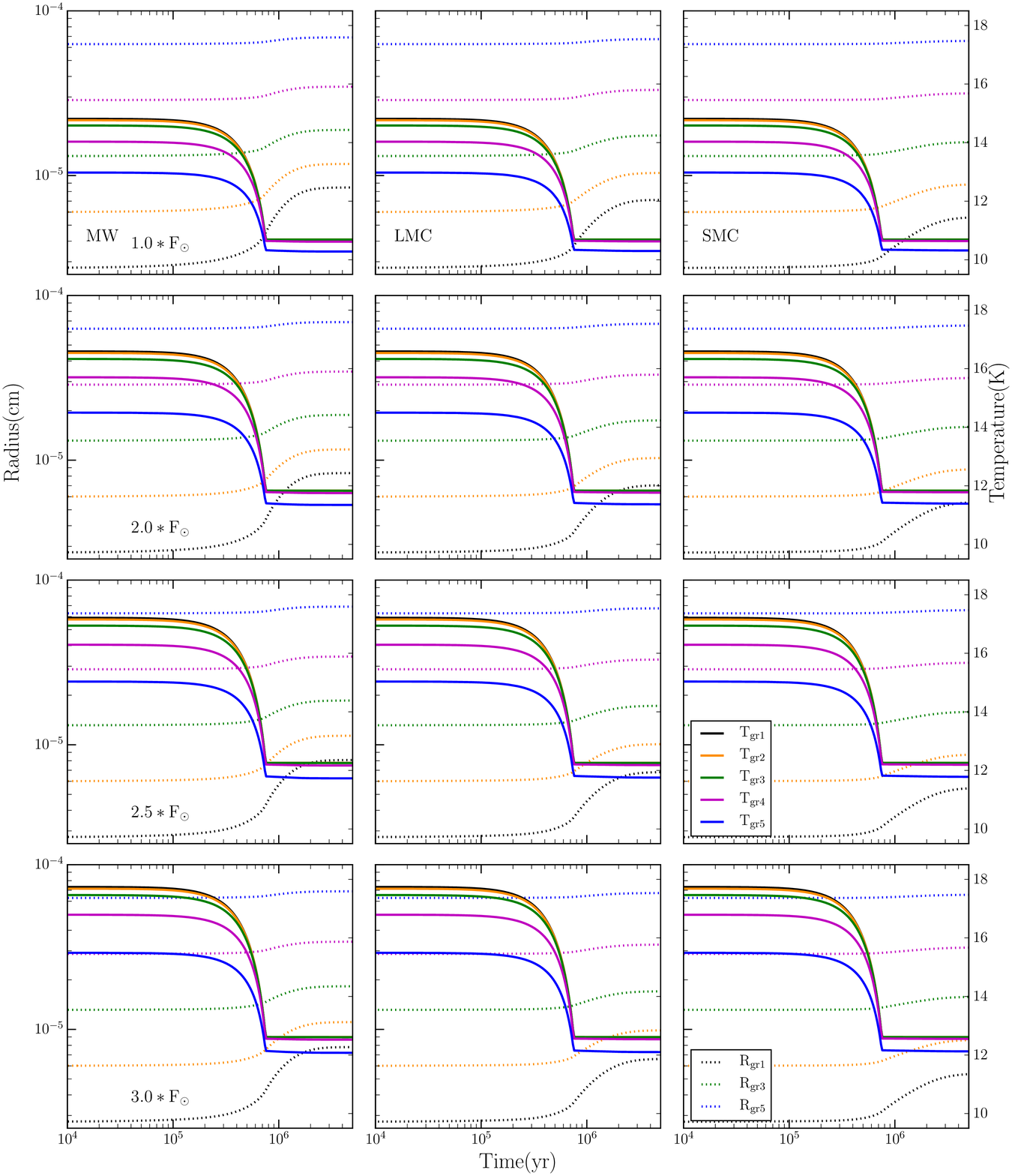}
\caption{Dust temperature (solid lines) and dust radius (dashed lines) versus time for the five grain sizes in the twelve models. Models with MW abundances are shown in the first column, LMC in the second, and SMC in the right column. Dust radius comprises the combined radius of the underlying dust grain plus the ice mantle. The first row has the base ISRF; ISRF increases with decreasing row with values of 2.0, 2.5 and 3.0.\label{fig:td_rd}
}
\end{figure*}

\begin{figure*}
\epsscale{1.17}
\plotone{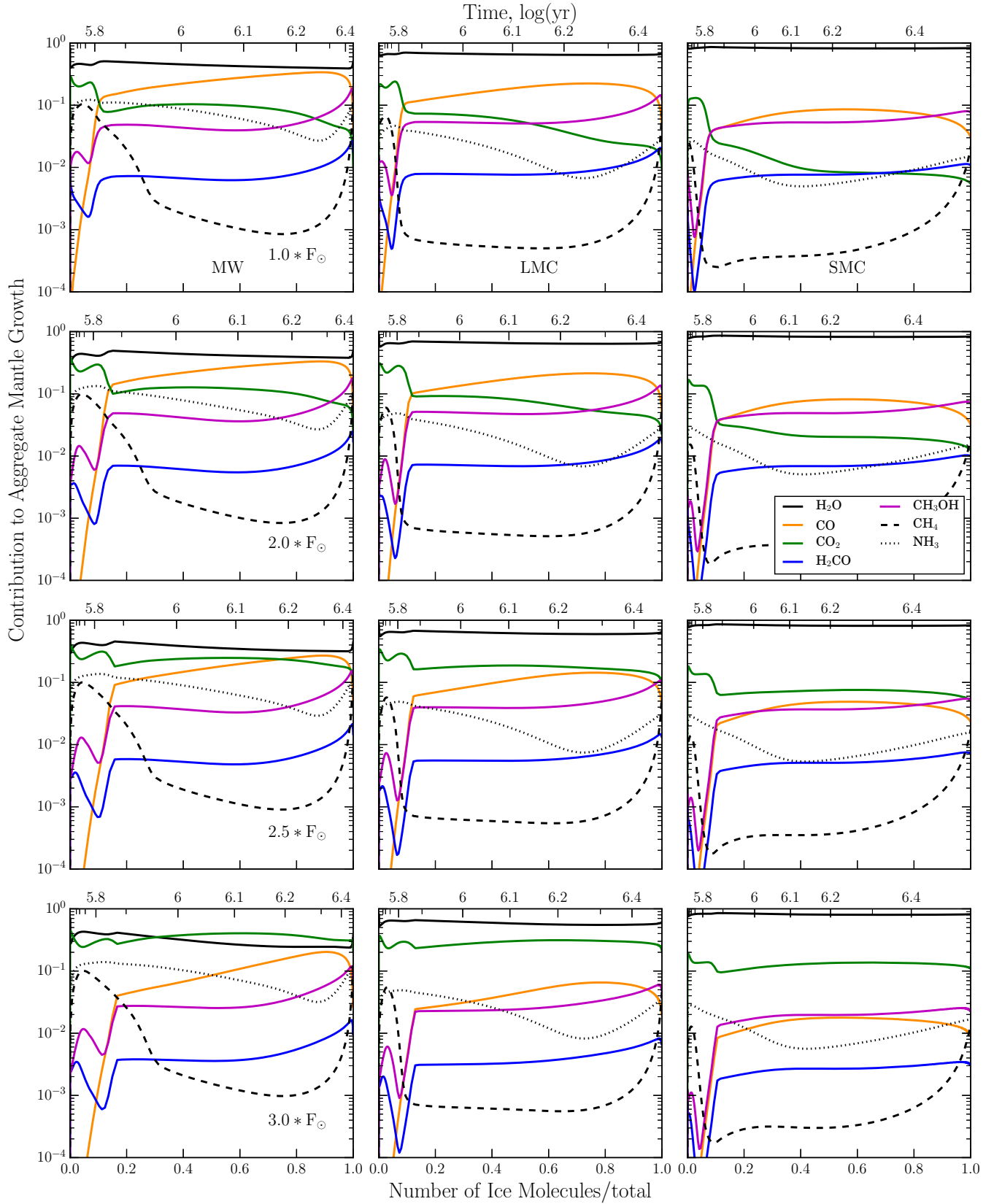}
\caption{Plots of the fractional surface layer abundance plotted against the growth of the aggregate ice mantle. For a given point in the accretion history of the ice mantle, the composition of newly formed mantle material can be read from the surface abundances, aggregated across all grain size bins. Model results with ISRF values of 1.5 are omitted.\label{fig:gr0}
}
\end{figure*}

\begin{table*}[ht!]
\begin{center}
\begin{tabular}{l l l l l l}
\toprule
Model & Grain 1 & Grain 2 & Grain 3 & Grain 4 & Grain 5\\
\midrule
Initial Radii & 0.0275 & 0.0601 & 0.1313 & 0.2872 & 0.6279 \\
\midrule
1.0\_MW & 0.0857 & 0.1185 & 0.1900 & 0.3471 & 0.6893 \\
1.5\_MW & 0.0853 & 0.1182 & 0.1898 & 0.3464 & 0.6888 \\
2.0\_MW & 0.0846 & 0.1174 & 0.1891 & 0.3458 & 0.6888 \\
2.5\_MW & 0.0821 & 0.1149 & 0.1868 & 0.3443 & 0.6894 \\
3.0\_MW & 0.0795 & 0.1122 & 0.1843 & 0.3420 & 0.6890 \vspace{0.2cm}
\\
1.0\_LMC & 0.0712 & 0.1038 & 0.1751 & 0.3312 & 0.6734 \\
1.5\_LMC & 0.0709 & 0.1036 & 0.1749 & 0.3309 & 0.6729 \\
2.0\_LMC & 0.0704 & 0.1030 & 0.1744 & 0.3305 & 0.6728 \\
2.5\_LMC & 0.0682 & 0.1009 & 0.1723 & 0.3290 & 0.6731 \\
3.0\_LMC & 0.0661 & 0.0987 & 0.1702 & 0.3270 & 0.6723 \vspace{0.2cm}
\\
1.0\_SMC & 0.0556 & 0.0882 & 0.1595 & 0.3154 & 0.6566 \\
1.5\_SMC & 0.0555 & 0.0881 & 0.1594 & 0.3153 & 0.6563 \\
2.0\_SMC & 0.0552 & 0.0878 & 0.1592 & 0.3151 & 0.6562 \\
2.5\_SMC & 0.0544 & 0.0870 & 0.1583 & 0.3145 & 0.6563 \\
3.0\_SMC & 0.0535 & 0.0861 & 0.1574 & 0.3136 & 0.6560 \vspace{0.2cm}
\\
1.0\_SMC\_gdr500 & 0.0773 & 0.1100 & 0.1813 & 0.3372 & 0.6786 \\ 
\bottomrule
\end{tabular}
\caption{Initial and final grain plus mantle radii, in $\mathrm{\mu m}$, to accompany Figure~\ref{fig:td_rd}. Note that all models begin with the same grain size distribution, given in the first line.\label{table:rd}}
\end{center}
\end{table*}

We computed a grid of fifteen models by multiplying the stellar component of the ISRF with values of [1.0, 1.5, 2.0, 2.5, 3.0] and varying the elemental abundances between initial elemental abundance setups, MW, LMC and SMC. Figure~\ref{fig:td_rd} shows the evolution of the dust temperatures and radii for twelve of the fifteen models. The end of collapse is apparent at $\sim 8 \times 10^5$ years, seen in the dust temperature minima; the model is then held at the final collapse density of 2 $\times$ 10$^4$ cm$^{-3}$ until $5 \times 10^6$ years. The visual extinction remains constant after the final density is reached, effectively fixing the dust temperatures; there is a slight decrease during this phase due to grain mantle growth, but the temperature-radius relation is roughly flat at $\mathrm{A_V \approx 10}$. At the final visual extinction of 10.6, the dust temperatures are equal for all but the largest grains, which are slightly cooler. Ice mantle growth primarily occurs at times immediately before and after the peak density is reached, when accretion onto the grains from the gas phase becomes rapid. The cross-sectional surface area is concentrated in the grains with small radius, causing the accretion rate to be highest for the smallest grains. The radius of this bin increases by up to a factor of three in models with high metal abundances; combined grain and ice radius values are given in Table~\ref{table:rd}. 

Figure~\ref{fig:gr0} shows, for a selection of models, the fractional ice-mantle composition by species, aggregated over all grain populations and plotted against ice layer depth. This ice depth is normalized to the final total ice abundance. Aggregate abundances for a given species are computed by first determining its fractional surface coverage on each grain size. Next, these fractional coverages are weighted by each grain's relative growth rate with respect to the total grain surface growth rate. These panels plot this weighted aggregate surface composition against the total ice abundance; the upper axis plots time for comparison. The mantle deposition rate for a given species depends directly upon its relative surface population, such that the surface composition is indicative of the newly-formed mantle composition at each point in model time. Therefore, these plots can be read as the mantle composition as a function of aggregate `layer'.

H$_2$O is the dominant ice component in nearly all models as expected, following observations. The collapse is complete by $\sim 8 \times 10^5$ ($\mathrm{10^{5.9}}$) years; for models with 1.0, 1.5 or 2.0 stellar intensity, this causes dust temperatures to drop below an efficiency threshold for producing CO$_2$ from CO + OH, identified by \citet{Garrod11}. CO mobility on the grain surface is sufficiently slowed at temperatures below $\sim$ 12 K; by this point, the fractional abundance of CO grows above that of CO$_2$. Models with 2.5 or 3.0 stellar intensity never drop below this temperature threshold, and as a result high CO$_2$ ice abundances are found throughout those models.

CH$_3$OH ice is formed via the hydrogenation of surface CO, which is only present after temperatures drop below the 12 K threshold. For MW models with abundant CO ice, the efficiency of CH$_3$OH formation appears low, with the abundance ratio of CH$_3$OH:CO ranging from 1:2 to 1:5. However, in SMC models with low CO surface abundance, surface CH$_3$OH can be equal in abundance to CO, and these molecules are similarly abundant throughout those model runs.

The hydrides CH$_4$ and NH$_3$ appear to track closely the elemental abundance of their atomic parent, with some dependence on temperature shown for models with galactic elemental abundances.

\begin{table*}[h!]
\begin{center}
\begin{tabular}{l l l l l l l l l l l l l l l }
\toprule
\multicolumn{1}{c}{Model} & \multicolumn{2}{c}{H$_2$O} & \multicolumn{2}{c}{CO} & \multicolumn{2}{c}{CO$_2$} & \multicolumn{2}{c}{CH$_3$OH} & \multicolumn{2}{c}{CH$_4$} & \multicolumn{2}{c}{NH$_3$} \\
\cmidrule(r){1-1}  \cmidrule(lr){2-3}  \cmidrule(lr){4-5}  \cmidrule(lr){6-7}  \cmidrule(lr){8-9}  \cmidrule(lr){10-11}  \cmidrule(l){12-13}
Time & 10$^6$ yr & $5 \times 10^6$ yr & 10$^6$ & $5 \times 10^6$& 10$^6$ & $5 \times 10^6$& 10$^6$ & $5 \times 10^6$& 10$^6$ & $5 \times 10^6$& 10$^6$ & $5 \times 10^6$ \\
\hline
1.0\_MW & 6.41(-5) & 1.51(-4) & 29.0 & 51.4 & 26.0 & 22.1 &  8.3  & 11.3 &  6.9 &  3.7  & 22.1 & 16.9 \\
1.5\_MW & 6.03(-5) & 1.47(-4) & 28.5 & 52.6 & 30.5 & 23.7 &  7.9  & 11.0 &  6.7 &  3.5  & 21.9 & 16.1 \\
2.0\_MW & 5.65(-5) & 1.39(-4) & 26.2 & 51.4 & 37.4 & 30.5 &  7.4  & 10.7 &  6.9 &  3.6  & 22.7 & 16.7 \\
2.5\_MW & 4.86(-5) & 1.12(-4) & 17.8 & 42.6 & 57.4 & 62.2 &  6.5  & 10.8 &  7.8 &  4.2  & 26.2 & 20.7 \\
3.0\_MW & 4.15(-5) & 8.51(-5) &  8.1 & 29.7 & 82.1 & 113.0 &  4.8  & 10.1 & 8.9 &  5.3  & 30.4 & 27.2 \\
\hline
1.0\_LMC & 4.28(-5) & 1.38(-4) & 14.2 & 24.7 & 16.2 &  9.3 &  6.2  &  8.8 &  1.6 &  0.8  &  5.3 &  3.1 \\
1.5\_LMC & 4.14(-5) & 1.37(-4) & 13.2 & 24.7 & 18.9 & 10.2 &  5.7  &  8.4 &  1.4 &  0.8  &  5.4 &  3.1 \\
2.0\_LMC & 4.00(-5) & 1.33(-4) & 11.6 & 23.1 & 22.3 & 13.7 &  5.2  &  8.1 &  1.4 &  0.7  &  5.5 &  3.2 \\
2.5\_LMC & 3.67(-5) & 1.17(-4) &  6.8 & 15.4 & 31.8 & 29.1 &  4.0  &  6.8 &  1.5 &  0.8  &  5.9 &  3.6 \\
3.0\_LMC & 3.40(-5) & 1.01(-4) &  2.6 &  6.8 & 41.2 & 48.6 &  2.3  &  4.5 &  1.5 &  0.8  &  6.3 &  4.1 \\
\hline
1.0\_SMC & 2.13(-5) & 8.55(-5) &  4.3 &  7.8 &  5.8 &  2.3 &  4.0  &  6.4 &  0.3 &  0.2  &  1.8 &  1.2 \\
1.5\_SMC & 2.10(-5) & 8.52(-5) &  4.0 &  7.8 &  6.9 &  2.6 &  3.6  &  6.2 &  0.2 &  0.2  &  1.8 &  1.2 \\
2.0\_SMC & 2.07(-5) & 8.41(-5) &  3.4 &  7.3 &  8.2 &  3.8 &  3.2  &  5.8 &  0.2 &  0.2  &  1.8 &  1.2 \\
2.5\_SMC & 2.00(-5) & 7.98(-5) &  2.0 &  4.4 & 11.3 &  9.3 &  2.3  &  4.3 &  0.2 &  0.2  &  1.9 &  1.2 \\
3.0\_SMC & 1.94(-5) & 7.53(-5) &  0.8 &  1.6 & 14.3 & 15.6 &  1.1  &  2.2 &  0.2 &  0.2  &  1.9 &  1.3 \\
\hline
1.0\_SMC\_gdr500 & 8.35(-6) & 7.70(-5) &  3.7 &  7.4 &  5.4 &  1.4 &  4.1  &  7.1 &  0.3 &  0.1  &  1.8 &  0.9 \\ 
\bottomrule
\end{tabular}
\caption{Fractional ice mantle abundances, shown for two times for each species and model; the left sub-column shows abundance (or fractional abundance with respect to water, in percent) at 10$^6$ years, while the right sub-column shows abundances at 5 $\times$ 10$^6$ years.\label{table:icevals}}
\end{center}
\end{table*}

\begin{table*}[h!]
{\small
\begin{center}
\begin{tabular}{l l l l l l l l l l l l l}
\toprule
\multicolumn{1}{l}{Model} & \multicolumn{2}{l}{H$_2$O} & \multicolumn{2}{l}{CO} & \multicolumn{2}{l}{CO$_2$} & \multicolumn{2}{l}{CH$_3$OH} & \multicolumn{2}{l}{CH$_4$} & \multicolumn{2}{l}{NH$_3$} \\
\cmidrule(r){1-1}  \cmidrule(lr){2-3}  \cmidrule(lr){4-5}  \cmidrule(lr){6-7}  \cmidrule(lr){8-9}  \cmidrule(lr){10-11}  \cmidrule(l){12-13}
Time & 10$^6$ yr & $5 \times 10^6$ & 10$^6$ & $5 \times 10^6$& 10$^6$ & $5 \times 10^6$& 10$^6$ & $5 \times 10^6$& 10$^6$ & $5 \times 10^6$& 10$^6$ & $5 \times 10^6$ \\
\midrule
1.0\_MW & 6.41(-5) & 1.51(-4) & 1.86(-5) & 7.74(-5) & 1.67(-5) & 3.33(-5) & 5.30(-6)  & 1.70(-5) &  4.44(-6) & 5.57(-6)  & 1.41(-5) & 2.54(-5) \\ 
1.5\_MW & 6.03(-5) & 1.47(-4) & 1.72(-5) & 7.73(-5) & 1.84(-5) & 3.49(-5) & 4.76(-6)  & 1.61(-5) &  4.01(-6) & 5.12(-6)  & 1.32(-5) & 2.37(-5) \\ 
2.0\_MW   & 5.65(-5) & 1.39(-4) & 1.48(-5) & 7.14(-5) & 2.11(-5) & 4.24(-5) & 4.19(-6)  & 1.49(-5) &  3.87(-6) & 4.94(-6)  & 1.28(-5) & 2.32(-5) \\ 
2.5\_MW & 4.86(-5) & 1.12(-4) & 8.63(-6) & 4.77(-5) & 2.79(-5) & 6.95(-5) & 3.18(-6)  & 1.21(-5) &  3.77(-6) & 4.74(-6)  & 1.27(-5) & 2.32(-5) \\ 
3.0\_MW   & 4.15(-5) & 8.51(-5) & 3.35(-6) & 2.53(-5) & 3.41(-5) & 9.62(-5) & 2.00(-6)  & 8.63(-6) &  3.69(-6) & 4.54(-6)  & 1.26(-5) & 2.31(-5) \\ 
\hline
1.0\_LMC & 4.28(-5) & 1.38(-4) & 6.08(-6) & 3.42(-5) & 6.93(-6) & 1.29(-5) & 2.64(-6)  & 1.22(-5) &  6.71(-7) & 1.12(-6)  & 2.27(-6) & 4.34(-6) \\ 
1.5\_LMC & 4.14(-5) & 1.37(-4) & 5.48(-6) & 3.38(-5) & 7.82(-6) & 1.41(-5) & 2.36(-6)  & 1.15(-5) &  5.98(-7) & 1.03(-6)  & 2.23(-6) & 4.28(-6) \\ 
2.0\_LMC & 4.00(-5) & 1.33(-4) & 4.64(-6) & 3.07(-5) & 8.93(-6) & 1.82(-5) & 2.08(-6)  & 1.07(-5) & 5.71(-7) & 9.87(-7)  & 2.20(-6) & 4.25(-6) \\ 
2.5\_LMC & 3.67(-5) & 1.17(-4) & 2.51(-6) & 1.81(-5) & 1.17(-5) & 3.40(-5) & 1.46(-6)  & 8.00(-6) &  5.43(-7) & 8.91(-7)  & 2.18(-6) & 4.22(-6) \\ 
3.0\_LMC & 3.40(-5) & 1.01(-4) & 8.96(-7) & 6.85(-6) & 1.40(-5) & 4.93(-5) & 7.91(-7)  & 4.53(-6) & 5.12(-7) & 7.83(-7)  & 2.15(-6) & 4.19(-6) \\ 
\hline
1.0\_SMC & 2.13(-5) & 8.55(-5) & 9.13(-7) & 6.68(-6) & 1.24(-6) & 1.93(-6) & 8.54(-7)  & 5.47(-6) &  5.95(-8) & 1.99(-7)  & 3.82(-7) & 1.00(-6) \\ 
1.5\_SMC & 2.10(-5) & 8.52(-5) & 8.29(-7) & 6.68(-6) & 1.46(-6) & 2.21(-6) & 7.60(-7)  & 5.25(-6) &  4.93(-8) & 1.86(-7)  & 3.79(-7) & 9.97(-7) \\ 
2.0\_SMC & 2.07(-5) & 8.41(-5) & 7.11(-7) & 6.13(-6) & 1.70(-6) & 3.22(-6) & 6.63(-7)  & 4.85(-6) & 4.36(-8) & 1.75(-7)  & 3.77(-7) & 9.95(-7) \\ 
2.5\_SMC & 2.00(-5) & 7.98(-5) & 4.02(-7) & 3.55(-6) & 2.26(-6) & 7.42(-6) & 4.53(-7)  & 3.44(-6) &  3.84(-8) & 1.49(-7)  & 3.75(-7) & 9.92(-7) \\ 
3.0\_SMC & 1.94(-5) & 7.53(-5) & 1.52(-7) & 1.24(-6) & 2.79(-6) & 1.18(-5) & 2.23(-7)  & 1.66(-6) & 3.37(-8) & 1.18(-7)  & 3.72(-7) & 9.88(-7) \\ 
\hline
1.0\_SMC\_gdr500 & 8.35(-6) & 7.70(-5) & 3.09(-7) & 5.70(-6) & 4.48(-7) & 1.07(-6) & 3.43(-7)  & 5.50(-6) & 2.16(-8) & 7.19(-8)  & 1.52(-7) & 6.75(-7) \\ 
\bottomrule
\end{tabular}
\caption{ The same data as from Table~\ref{table:icevals}, but given as absolute values rather than fractional abundances with respect to H$_2$O. Abundances are shown for two times for each species and model; the left sub-column shows abundance at 10$^6$ years, while the right sub-column shows abundances at 5 $\times$ 10$^6$ years.\label{table:iceabsvals}}
\end{center}
}
\end{table*}

The following subsections describe the important reactions producing and destroying each primary ice component; we refer to relative abundance trends seen in Figure~\ref{fig:gr0} or to abundance values at $10^6$ or $5 \times 10^6$ years, found in Tables~\ref{table:icevals} and~\ref{table:iceabsvals}.

\subsection{H$_2$O Ice Behavior}
H$_2$O ice formation occurs primarily through the surface hydrogenation of OH, which is in turn formed via O + H on the surface. Prior to the completion of collapse, if dust temperatures are greater than $\sim$13.5 K, reaction proceeds primarily through OH + H. At these dust temperatures, desorption of H$_2$ is strongly competitive with the barrier-mediated OH + H$_2$. After collapse, the dust is cool enough such that H$_2$ resides on the grain surface for sufficient time to react and becomes the dominant H$_2$O formation route.

Figure~\ref{fig:gr0} shows that H$_2$O is the most abundant ice mantle component for all models except those with MW elemental abundances at high stellar intensity, 2.5\_MW and 3.0\_MW. In these models the CO gas abundance is high, and post-collapse temperatures are warm enough for CO mobility on the grain surface. These effects combine for CO + OH to compete effectively with H + OH and H$_2$ + OH, reducing H$_2$O ice abundance while enhancing CO$_2$. In models with reduced elemental abundances, CO never attains the surface coverage required for CO$_2$ production to reach similar levels. Additionally, the decreased C/O ratio in `LMC' and `SMC' chemistries further enhances H$_2$O dominance over carbon-bearing ice species.

The absolute abundance of H$_2$O ice (Table~\ref{table:icevals}) does not strictly follow the abundance of oxygen across the different models; because the carbon abundance serves to lock oxygen into CO-structured molecules, the fraction of oxygen found in H$_2$O is determined in large part by the C/O ratio. As dust temperature increases due to increased stellar intensity, H$_2$O ice abundance drops. This is due to increased competition between OH + CO and OH + H/H$_2$; increasing temperatures serves to increase the fraction of OH going towards CO$_2$ formation via increased CO mobility, while the production of H$_2$O decreases in turn.

\subsection{CO Ice Behavior}
CO is efficiently formed in the gas phase and accretes onto the grain surface. At early times in the model nearly all surface CO reacts with OH to form CO$_2$ due to mobile CO on the warm ($\gtrsim 12 K$) dust. If the dust temperature remains high after collapse, CO$_2$ efficiently forms at late times as well. At post-collapse densities, the hydrogenation of CO into the short-lived HCO also becomes an important process, with possible outcomes of reverting to CO or forming stable H$_2$CO. H$_2$CO can then be further hydrogenated to form methanol, CH$_3$OH. These reactions serve to destroy CO ice; however, if accretion rates are comparable to the rate of these destruction reactions, the transport of CO to the mantle phase can proceed before the destruction of all surface CO, leading to non-negligible CO mantle abundances and less efficient conversion to methanol. 

Figure~\ref{fig:gr0} shows low CO abundances pre-collapse for all models. Post-collapse, the behavior is determined by the dust temperature and the accretion rate. Accretion rate is driven by the amount of carbon, oxygen and other heavy elements in the model; it is lowest in SMC models and highest in MW models. Higher accretion rates will drive more surface CO into the inert mantle simply by building up the ice layers more quickly, leading to higher CO ice abundances in the mantles. The post-collapse dust temperature determines the efficiency of CO$_2$ formation; for models with less than 2.5 times the base stellar intensity, an inversion in the CO:CO$_2$ ratio is seen at the end of collapse, while a stronger interstellar radiation field allows strong CO$_2$ formation even to A$\mathrm{_V}$ of 10.6. This depletes CO levels for the entirety of the model.

\begin{figure*}
\label{fig:sepgr1p0}
\epsscale{1.15}
\plotone{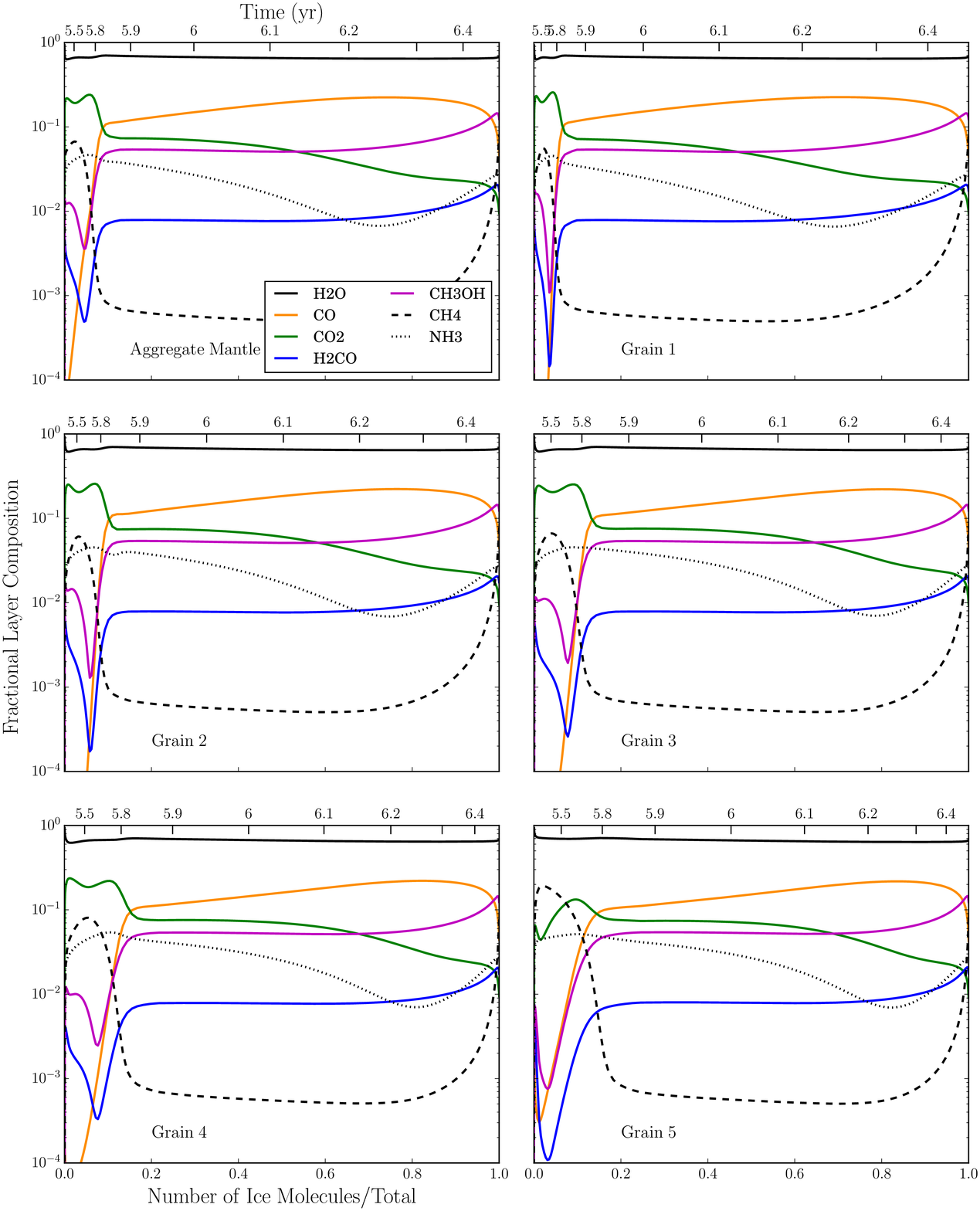}
\caption{These panels plot the fractional layer composition as a function of total mantle plus surface abundance (or time) for the model 1.0\_LMC. The aggregate layer composition is shown in the top left panel, as in Figure~\ref{fig:gr0}. The following five panels depict the fractional layer abundances for the five grain size populations in the model. Grain sizes for the numbered panels are given in Table~\ref{table:rd}, with `Grain 1' being the smallest grain size.}
\end{figure*}

\begin{figure*}
\label{fig:sepgr3p0}
\epsscale{1.15}
\plotone{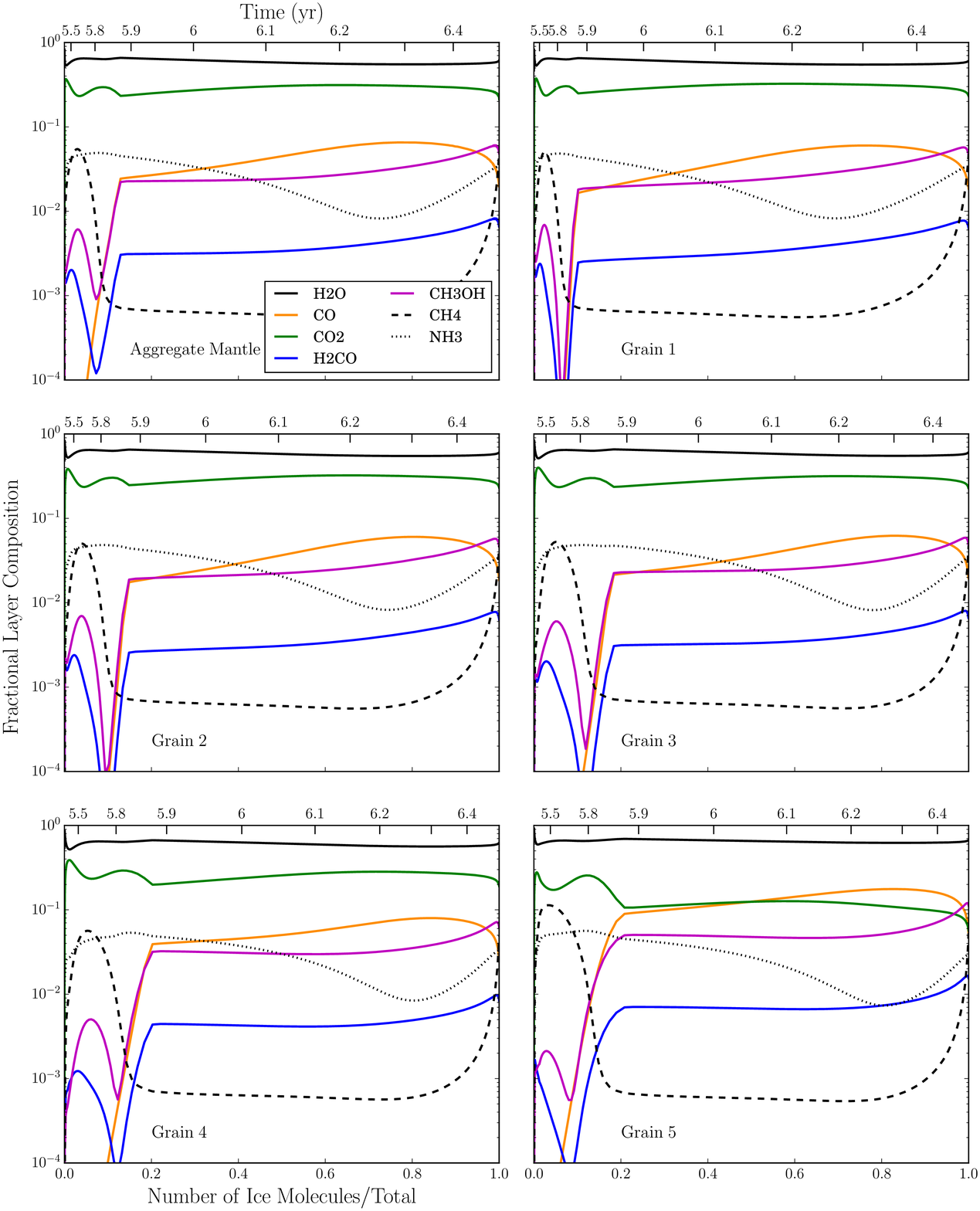}
\caption{Fractional layer compositions, as in Figure~\ref{fig:sepgr1p0}, but for the model 3.0\_LMC.}
\end{figure*}

The surface abundances of the five grain populations are shown separately in Figures~\ref{fig:sepgr1p0} and \ref{fig:sepgr3p0} for two models, 1.0\_LMC and 3.0\_LMC. In Figure~\ref{fig:sepgr1p0}, the abundance of CO is seen to increase dramatically as the model collapses and grain temperatures drop below the 12 K threshold. The exact time of the abundance turnover is different for the individual grain populations, as each has a different temperature. Figure~\ref{fig:sepgr3p0} shows a model dominated by CO$_2$ for all but the largest grains due to elevated dust temperatures induced by the elevated ISRF value of 3.0. The relative drop in temperature is also more extreme, and the finite number of time steps is apparent in these plots due to a rapid transition in chemical behavior.

\subsection{CH$_3$OH Ice Behavior}

Hydrogenation of CO to CH$_3$OH has two steps with activation energy barriers that produce short-lived radicals, HCO and CH$_2$OH/CH$_3$O. Hydrogenation of HCO will form products of either H$_2$CO or H$_2$ + CO with equal probability, an assumption of our model. Once formed, H$_2$CO is fairly robust to reverting to a less-hydrogenated form, reacting with a hydrogen atom to form CH$_3$O or CH$_2$OH more readily than HCO + H$_2$. Hydrogen addition to H$_2$CO and abstraction from CH$_3$OH are fast, as manifested in the constant ratio of surface abundances between the two species across all models.

The total abundance of CH$_3$OH ice in the models shown in Table~\ref{table:iceabsvals} has little spread, with variation of only a factor of two to four across models with an order of magnitude less elemental carbon abundance (MW to SMC). Notably, the amount of CH$_3$OH relative to the amount of CO on the grain surface increases as the elemenatal carbon abundance decreases across models. The change is primarily driven by a strong decrease in CO ice abundance as elemental abundances decrease, from MW to LMC to SMC values. For a set of models with equal elemental abundances, the abundance of CH$_3$OH drops by a factor of two to four as the ISRF increases from 1.0 to 3.0, showing a decrease in formation efficiency at higher dust temperatures.

\subsection{CH$_4$ and NH$_3$ Ice Behavior}

These ices form primarily through successive hydrogenation on grain surfaces. NH$_3$ ice has a linear pathway with little branching, though N$_2$ can be a significant nitrogen carrier for models with high nitrogen abundance. CH$_4$ ice shows similar behavior; the primary formation of CH$_4$ begins with atomic carbon. The sharp decline in CH$_4$ ice abundance shown in Figure~\ref{fig:gr0} at early time is indicative of carbon forming CO in the gas phase and the atomic abundance decreasing rapidly. Because nitrogen has no equivalent reservoir, its ice behavior is more consistent throughout mantle formation.

The total abundance of NH$_3$ ice shown in Table~\ref{table:iceabsvals} reflects a consistent fraction of total nitrogen found in NH$_3$ ice across models with varying elemental nitrogen abundance. However, CH$_4$ ice does not follow this trend, with elevated abundance of CH$_4$ per carbon atom in models with increased elemental carbon abundance. This reflects the increasing C/O ratio in models with increasing carbon abundance. Models with increased C/O ratio take longer to convert gas-phase atomic carbon into CO; as the formation of surface hydrocarbons requires accretion of atomic carbon, models that sustain a reservoir of atomic carbon in the gas show elevated CH$_4$ abundances.

Figure~\ref{fig:sepgr1p0} shows the ice compositions of individual grain sizes in the distribution, specifically for the 1.0\_LMC model. CH$_4$ behavior on the largest grain size differs from its counterparts at early times ($\sim$ 3 $\times$ 10$^5$ years). CH$_4$ surface abundance is greater than CO$_2$ for the largest grain; this is caused primarily by the difference in temperature between the grain sizes, with lower temperatures enhancing hydrogenation rates of atomic carbon. The destruction of atomic carbon on large grains is almost entirely through C + H $\rightarrow$ CH, while on small grains roughly 20\% of carbon reacts via C + OH $\rightarrow$ CO + H. Additionally, the warmer temperatures on small grains permits diffusion of the CH$_3$ radical, opening new pathways for destruction of CH$_3$ via e.g. CH$_3$ + CH$_3$ $\rightarrow$ C$_2$H$_6$, a molecule not yet detected in interstellar regions but strongly detected in the coma of Comet Hyakutake \citep{Mumma96}. Ethane surface and mantle abundances are highest on the small grains, with a C$_2$H$_6$:CH$_4$ grain abundance ratio in the model 1.0\_LMC of 1.0 at 3 $\times$ 10$^5$ years, dropping to 0.1 at 10$^6$ years.

\subsection{Gas to Dust Ratio}
The gas to dust ratio was fixed at a value of 175 throughout the model grid. This is known to vary with environment, but we chose to keep it fixed to disentangle its effects from the effects of changing elemental abundances and dust temperatures. We ran an additional model with an ISRF value of 1.0 and SMC abundances at a gas to dust ratio of 500 to test the robustness of the grid results.

The aggregate mantle of the reduced dust model is shown in Figure~\ref{fig:gdr500}, below the equivalent grid model with a ratio of 175; absolute ice abundances are given in Table~\ref{table:iceabsvals}. The decreased aggregate cross-sectional area lowers the total accretion rate. This serves to lower CO$_2$ abundance, which requires accretion of CO during the warm pre-collapse phase. Other species are comparable between the two models, due to the long post-collapse phase from 10$^6$ to 5$\times$10$^6$ years. The comparable abundances of solid species cause mantles to be appreciably thicker in the model with gas to dust ratio set at 500, seen in Table~\ref{table:rd}.

Notably, the model with an increased gas to dust ratio also exhibits enhanced CH$_3$OH abundance. The increased formation efficiency of CH$_3$OH appears to be directly connected to the heavy atom accretion rate onto grain surfaces, the key driver of inert ice mantle growth. If the ice mantle does not grow quickly enough to sequester CO, long surface times coupled with high hydrogen accretion rates produce a high CH$_3$OH abundance.

\section{Discussion}

\begin{figure}
\label{fig:compare}
\includegraphics[width=\columnwidth]{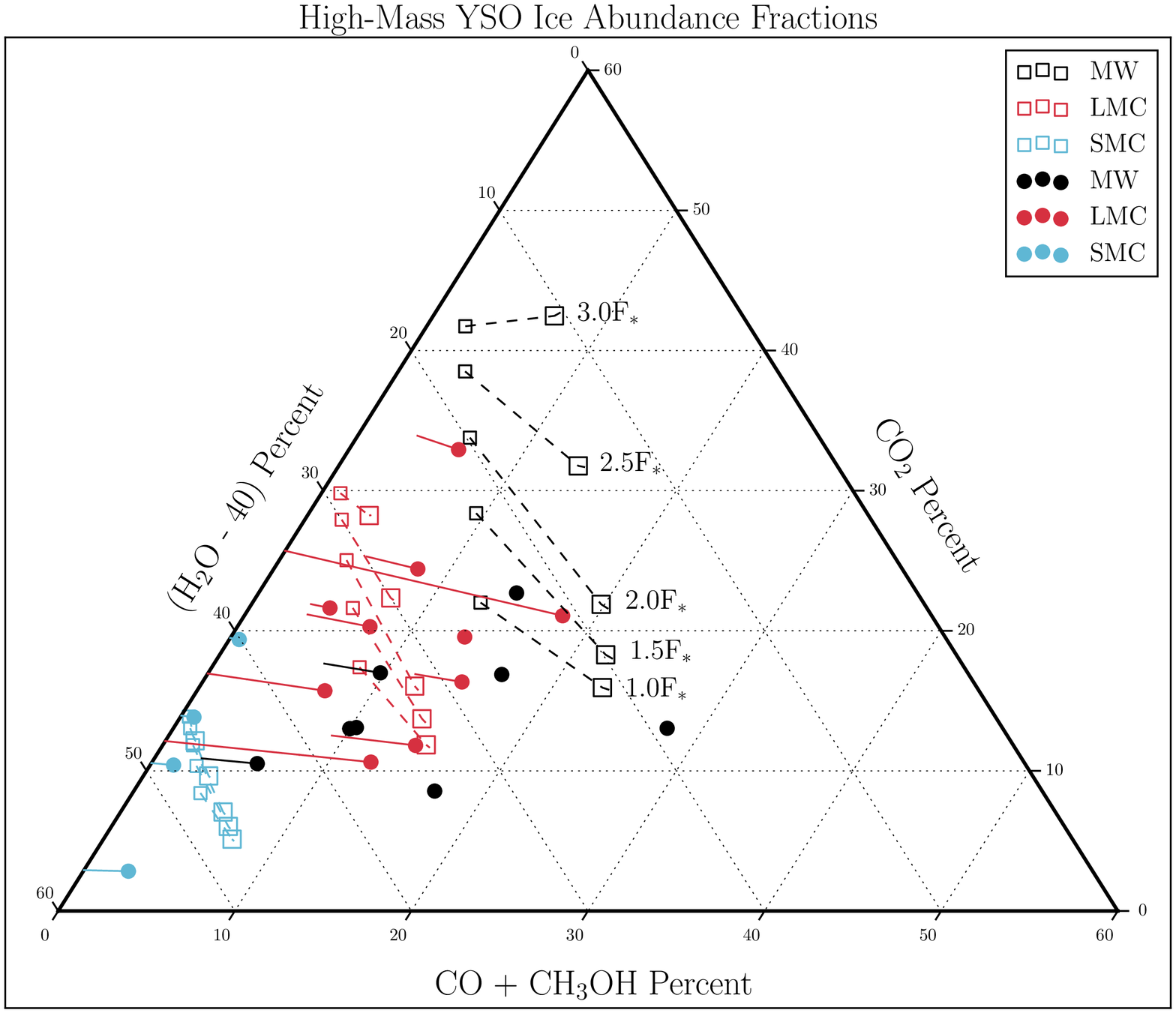}
\includegraphics[width=\columnwidth]{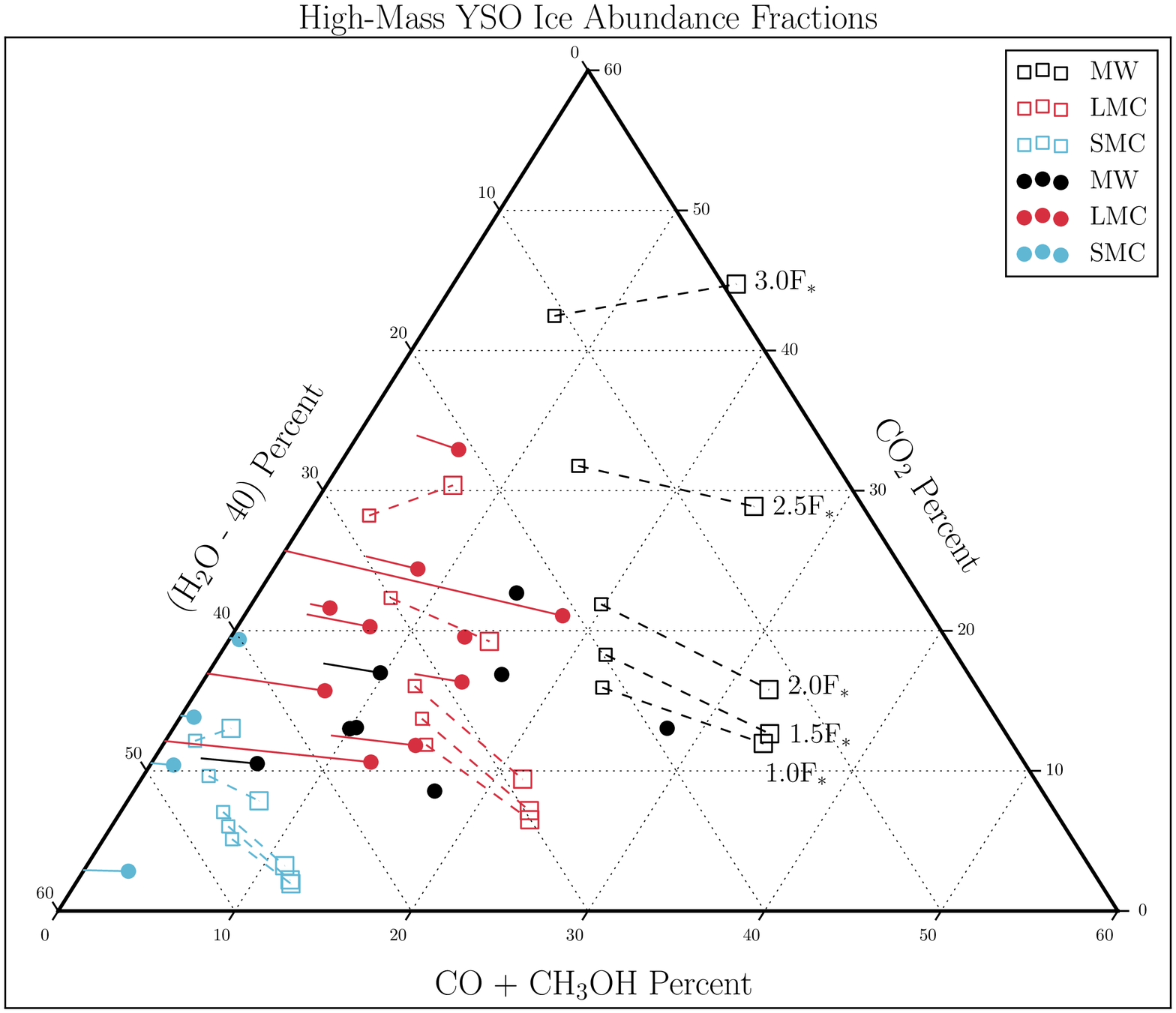}
\caption{These panels overlay the model results on the observations from Figure~\ref{fig:obs}. With matching colors for model elemental abundances to source environment (MW, LMC, SMC), models with increasing stellar flux parameter move upwards on the plot, with galactic chemical abundance models labeled. For each model, two points are plotted; for the top panel, the smaller leftward square shows the ice composition at the time of collapse completion ($\sim$8$\times$10$^5$ years), while the larger rightward square shows the composition at 10$^6$ years. For the lower panel, the leftward square shows the composition at 10$^6$ years, while the rightward square shows the values for 5$\times$10$^6$ years. \citep[Ternary figure style from][]{Harper15}}
\end{figure}

Figure~\ref{fig:compare} shows the model results overlaid on the observations. For each model we plot two points per panel; for the top panel, the smaller leftward point shows the ice composition at 8$\times$10$^5$ years, while the large rightward point shows the composition at 10$^6$ years. The lower panel shows the same models at times of 10$^6$ years and 5$\times$10$^6$ years. The choice of time spent at post-collapse density is arbitrary, and because physical conditions do not change significantly during this time, the composition follows a roughly straight line between these points. The MW and SMC abundance models appear to match observations more closely at earlier times, while the LMC observations have variation such that model matches are found at both early and late times.

Models with MW abundances show compositions enriched in CO$_2$ and CO/CH$_3$OH, with the relative enrichment between the two set by the stellar flux parameter. As no observations lie near the high-flux galactic abundance models, these models do not appear to represent any observed MYSO and can be ignored.

LMC MYSOs demonstrate large variation in composition, from extreme CO/CH$_3$OH depletion (as in SMC sources) to highly enriched in CO$_2$. Models with LMC-like elemental abundances fall across the ensemble of LMC MYSOs, though there is considerable overlap between LMC and galactic sources. The models separate cleanly on this plot, implying additional effects not addressed in the model setup. Variation in local metallicity may cause blending, as metal-poor MW YSOs may appear chemically similar to LMC MYSOs. Parameters beyond our model, such as variation in collapse speed or ice processing may also play a role.  Models are able to fit LMC observations at the full range of stellar flux parameter tested.

Models with the most depleted elemental abundances fall near the observed SMC MYSOs, matching the low (undetected) CO abundance and presence of CO$_2$. The models lying closest to observed YSO abundances have high stellar flux values, though the models cannot fully reproduce the spread in CO$_2$ abundances and typically overproduce CO/CH$_3$OH. Of note, the composition of SMC models in Figure~\ref{fig:gr0} shows a roughly equal abundance of surface CO and CH$_3$OH, while observational upper limits exist only for CO. Tightening the abundance constraints on these two species would provide strong evidence for the validity of our CO surface chemistry. 

The increased CH$_3$OH abundance relative to CO in SMC models is an unexpected result. CH$_3$OH formation requires CO surface residence times to be longer than the mantle deposition timescale to allow sufficient time for hydrogenation. In this way, the balance between CO and CH$_3$OH is determined primarily by the accretion rate of elements heavier than hydrogen. With long CO surface residence times, CO$_2$ production will also be increased if dust temperatures are above the necessary threshold for CO surface mobility.

As discussed by~\citet{Garrod11}, the main reaction that forms grain-surface CO$_2$, i.e. CO + OH $\rightarrow$ CO$_2$ + H, involves internal barriers whose behavior in the gas phase may be approximated by a single, modest (80~K) barrier. On grain surfaces, this barrier is much lower than the barrier to the diffusion of either CO or OH, meaning that, upon meeting, these reactants will usually be confined together long enough for the reaction to take place, giving the reaction an effective efficiency close to unity.

\citet{Garrod11} determined a minimum temperature of around 12 K for CO to be efficiently converted to CO$_2$ on grains, corresponding to the temperature at which CO becomes sufficiently mobile on the grain surface not only to be able to meet its reaction partner OH, but for the rate of reaction of CO + OH to be able to compete effectively with the reaction H$_2$ + OH $\rightarrow$ H$_2$O + H, which in this model has an activation energy barrier of 2100~K, again based on gas-phase estimates, following~\citet{Garrod11}. For the latter reaction, in spite of its ability to occur through tunneling, the diffusion of H$_2$ away from its reaction partner, OH, is nevertheless more probable than reaction. This means that tunneling through the reaction barrier is the rate-limiting step in the reaction, and not the diffusion of H$_2$. Consequently, variation of the diffusion rate of H$_2$ with temperature does not affect either the rate of the H$_2$ + OH reaction, nor the ability of the CO + OH reaction to compete with it. Variations in temperature do, however, affect the average population of H$_2$ on the grain surfaces, which is determined almost entirely by the balance between the rate of accretion of gas-phase H$_2$ molecules onto the grains and the rate at which they thermally desorb back into the gas. The variation in the H$_2$ population with temperature has a direct effect on the competition between the CO + OH and H$_2$ + OH reactions, with higher temperatures acting to reinforce the dominance of CO, due to the reduction in H$_2$ population.

The suggested threshold temperature of 12~K for CO to CO$_2$ conversion is necessarily approximate, as it is representative of a range of temperatures for which CO$_2$ conversion may range from close to 100\% down to a low-temperature conversion ratio somewhere on the order of 10\% (or less). As discussed above, the threshold will be somewhat dependent on the rate of accretion of H$_2$ onto the grains, which scales with gas density. The picture is further complicated by the effects specifically studied in this paper, in which a range of grain sizes and temperatures contribute to an aggregate composition, in some cases including grains that fall above and below the threshold of efficient CO$_2$ production. For models presented in this paper, peak CO$_2$ surface formation is found on grains with temperatures in the range of roughly 14 to 18 K; we do not explore temperatures above 18 K, where thermal evaporation of species likely curtails CO$_2$ formation, while temperatures lower than 14 K allow for H$_2$ + OH to compete noticeably in the destruction of surface OH.

However, we may consider how much variation there could be in the guideline threshold temperature through various influences. The discussion in the above paragraphs demonstrates that the main determinants of this temperature are the diffusion rate of CO, the desorption rate of H$_2$, and the accretion rate of H$_2$. A simplistic consideration of the balance between the rates of these processes at a nominal temperature of 12~K indicates that an order of magnitude increase in gas density, increasing the accretion rate of H$_2$ by the same factor, would produce a commensurate increase in the threshold temperature of $\sim$0.8 K. The adoption of an H$_2$ binding energy 10\% smaller than the 430~K used in our model would increase the threshold temperature by 0.65~K. The use of a CO diffusion barrier equal to 30\% of the CO binding energy, rather than the 35\% we adopt here, would decrease the threshold temperature by $\sim$0.83~K. (The review by \citet{Cuppen17} suggest that this ratio for CO lies between 30--40\%.)

It should be noted that changing each of the above parameters in the opposite sense would produce a similar variation in the threshold temperature in the other direction. However, each of the determined variations has been calculated in isolation, and it is unclear how a combination of different parameters would affect the overall threshold. Under certain conditions, the models also fall into the so-called accretion limit, under which modified rate equations become active in the model \citep[see][]{Garrod08}. Such conditions will change the simple treatment of the balance between processes that we consider above. A robust determination of the sensitivity of the threshold temperature to each parameter demands a more rigorous testing of the parameter space using the full chemical/physical model. The continuing refinement of laboratory measurements of CO and H$_2$ binding and diffusion properties will also be very valuable to this effort.

\subsection{Thermal Ice Processing}
The models produce a reasonable fit to observations, though a general trend exists in overproduction of (CO + CH$_3$OH). This may not be a simple model issue but instead a comparison of model results to observations in different physical regimes. These MYSOs are highly luminous objects, and thermal processing of the envelope is likely to have occurred in many sources. In this case, the most volatile ices may be under-abundant due to evaporation when compared to the final model output, which ends prior to a grain heating and ice evaporation phase.

\citet{Collings03a,Collings03b} find via temperature - programmed desorption (TPD) studies that CO bound to a CO substrate desorbs at temperatures of $\sim$ 25 K. TPD of CO bound to an H$_2$O surface find desorption temperatures between 30-70 K, depending on the nature of the H$_2$O ice deposition. Residual CO ice is able to linger in the H$_2$O ice until $\sim$ 140 K, when H$_2$O ice crystallization causes so-called ``volcano-desorption''. If temperatures reach $\gtrsim$ 70 K, CO$_2$ will begin desorption from an H$_2$O surface \citep{Fayolle11,Noble12}. Entrapment of CO$_2$ in the H$_2$O ice will prevent complete removal of CO$_2$, though relative loss is dependent on ice thickness, mixing ratio, and other parameters. Complete loss is not expected until H$_2$O crystallization. The temperatures quoted here apply to laboratory timescales of minutes to hours, but fitting the desorption rates to an activation energy barrier via the Polanyi-Wigner equation provides a useful quantity for models with astrophysical timescales \citep[see e.g. the desorption of CO at roughly 25 K in][]{Garrod13}. 

These experimental results provide evidence for the observations having gone through some amount of mantle desorption; models with a complete collapse to high densities and a following warm-up phase may better account for this effect.

\subsection{Comparison to Polar \& Apolar Ice Features}
We have used a single-point model to trace the collapse of the envelope; this approach cannot fully reproduce the signature of envelope shells with varying age. The outer regions of the envelope may be chemically younger than the central source, and this difference in total ice formation between regions may be significant depending on the age of the outer envelope. Ice formed in this young environment would be CO$_2$-rich and CO-poor, while significant freeze-out in the inner high density region forms a significant surface abundance of nearly pure CO~\citep{Pontoppidan03}. The densities reached in our models are not sufficiently high to achieve such strong freeze-out of CO that it becomes dominant, although in the Milky Way models it comes close. Figure 12 of~\citet{Garrod11} shows a high density collapse in which CO eventually dominates H$_2$O. However, it is unclear whether apolar CO signatures require numerical dominance over the total surface H$_2$O abundance, or whether these signatures can be achieved with some fractional abundance of CO, combined with some surface self-segregation mechanism that is effective even at low temperatures.

\subsection{Cosmic Ray Ionization Rate}
The cosmic ray ionization rate, $\zeta$, was held fixed throughout our model grid at a value of 1.3 $\times 10^{-17}$ s$\mathrm{^{-1}}$. Other chemical model work in the Magellanic Clouds have used either the galactic local value or an enhanced value \citep{Chin98,Acharyya15,Acharyya16}. Data on $\zeta_{LMC}$ and $\zeta_{SMC}$ is scarce; \citet{Abdo10LMC} analyzed a Fermi-LAT \textgreater 100 MeV gamma ray map of the LMC and found the globally-averaged cosmic ray ionization rate to be 20-30\% of the local MW value. Regional variability can be significant, with cosmic ray sources in the LMC causing nearby regions to have ionization rates higher than the globally averaged value. SMC studies lack the sensitivity and resolution required for anything other than a global measurement; this value is depleted by at least a factor of six to seven with respect to the local galactic value \citep{Abdo10SMC}.

\begin{figure}[ht!]
\label{fig:gdr500}
  \begin{tabular}[b]{@{}p{\columnwidth}@{}}
    \centering\includegraphics[width=\columnwidth]{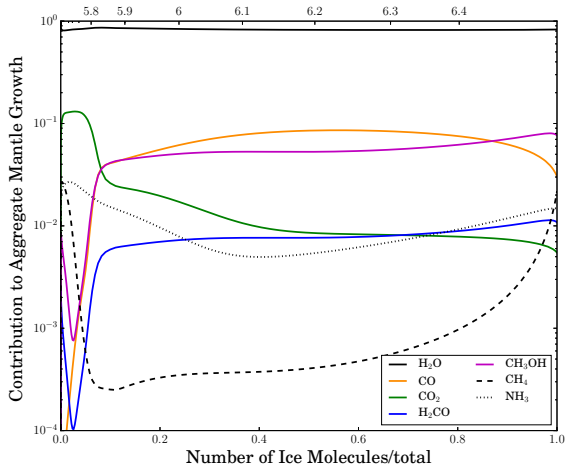} \\
    \centering\small (a) Gas to Dust Ratio: 175 \\
    \centering\includegraphics[width=\columnwidth]{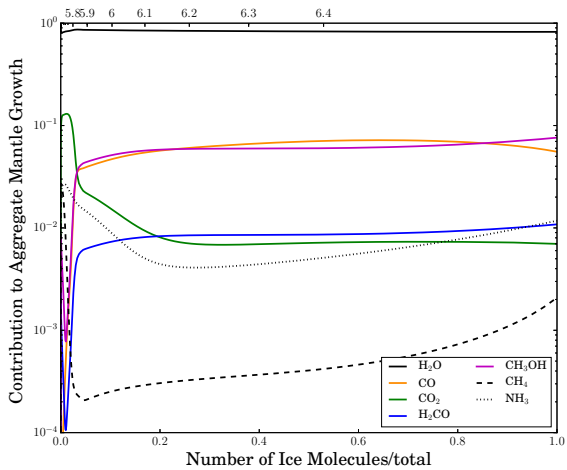} \\
    \centering\small (b) Gas to Dust Ratio: 500\\
  \end{tabular}
\caption{These panels compare models with 1.0 F$_*$ and SMC abundances but differing gas to dust ratios. Panel (a) shows the model with a ratio of 175 while panel (b) shows the model with a ratio of 500. }
\end{figure}

\subsection{Comparison to Low-Mass YSOs}
The model results presented here have been compared with high-mass YSOs due to the availability of such data for the Magellanic Clouds. However, low-mass YSOs in the Milky Way generally demonstrate a larger solid-phase CO$_2$/H$_2$O ratio than high-mass sources \citep[see e.g.][]{Oberg11}. Due to the simple free-fall collapse model we employ here, the models do not directly address differences between high- and low-mass sources. It is plausible that the dust temperature differences investigated here could be responsible for such variations, perhaps with the envelopes of lower-mass sources spending longer periods at the higher temperatures more conducive to CO$_2$ production. The inclusion of a more explicit temperature structure in the models would help to elucidate this issue. 

\section{Summary}
Our results suggest that gas-grain models of cold cloud collapse can produce ice mantle abundances that match reasonably well to observations in a variety of environments. We conclude that:
\begin{itemize}
\item The values of ISRF intensity and elemental abundances chosen provide an adequate distribution of ice abundances that cover the observed ice abundances in YSOs. Models with strongly enhanced ISRF intensity at MW elemental abundances are excluded, while SMC models with enhanced ISRF are preferred.
\item LMC models lie near observed YSOs for every value of the ISRF intensity modeled, characterizing the large spread in LMC YSO ice abundances. This may be indicative of large local fluctuations in the LMC ISRF.
\item The ISRF intensity strongly affects the relative abundance of CO$_2$ to CO/CH$_3$OH, with higher ISRF values leading to CO$_2$ enhancement. This is caused by a temperature threshold for CO mobility on grain surfaces, leading to efficient production of CO$_2$ at dust temperatures $\gtrsim$ 12 K.
\item Increasing model elemental abundances (and corresponding C/O ratio) decreases the H$_2$O abundance against the other ices; this is evidenced by model values moving parallel to the H$_2$O ternary axis with changes in elemental abundance.
\item Our models indicate that the lack of CO in SMC sources is most likely caused by a combination of low elemental abundances and high ISRF intensity.
\item CH$_3$OH abundance is found to be enhanced in low-metallicity environments relative to CO. The enhancement is caused by the relatively slow accretion rate in the low-metallicity models; CO is more efficiently hydrogenated due to longer surface residence time, and the production of CH$_3$OH increases. This is an important start for the formation of complex organic molecules in LMC and SMC hot cores. 
\end{itemize}

We leave some issues to be addressed in future work. Thermal processing of the ice is important for matching observed ice abundances, and it is not included in these models. We find significant growth in the [dust+mantle] radius, which affects both the dust temperature and surface chemistry; however, we assume a Q$\mathrm{_{abs}}$ of carbonaceous dust for temperature calculations, though the Q$\mathrm{_{abs}}$ of ice will differ. We also use a grain size distribution found for silicate grains; this could be resolved by using values for silicate or carbonaceous grains throughout, or by attempting to model both populations. 

Future models could investigate the dependence on cosmic ray ionization rate, a parameter with large variation across the LMC. The rate of collapse may also be important, as it sets the heavy atom accretion rate. Follow-up models will address behavior in collapse to higher densities ($\sim$ 10$^7$ cm$^{-3}$), including a warm-up phase for comparison to a newly detected hot core in the LMC \citep{Shimonishi16b}.

\acknowledgments
The authors would like to thank the anonymous referee for their helpful comments which greatly improved the paper. TP would like to thank Ilsa Cooke and Klaus Pontoppidan for informative conversations on ice experiments and observations. TP was supported by a NASA Earth and Space Sciences Fellowship.

\software{MAGICKAL \citep{Garrod13, Pauly16},
		Matplotlib \citep{matplotlib},
		Jupyter Notebooks \citep{jupyter}}

\bibliographystyle{yahapj}
\bibliography{refs}

\end{document}